\documentstyle[aps,prl,times,epsfig]{revtex}

\newcommand{\bx}{{\bf x}}

\newcommand{\bdx}{\delta {\bf x}}
\newcommand{\bmu}{\mbox{\boldmath $\mu$}}
\newcommand{\bpi}{\mbox{\boldmath $\pi$}}
\newcommand{\bc}{{\bf c}}
\newcommand{\bC}{{\bf C}}
\newcommand{\bv}{{\bf v}}
\newcommand{\bF}{{\bf F}}
\newcommand{\bX}{{\bf X}}

\newcommand{\tinit}{t_0}

\newcommand{\BE}{\begin{equation}}
\newcommand{\EE}{\end{equation}}
\newcommand{\BA}{\begin{eqnarray}}
\newcommand{\EA}{\end{eqnarray}}

\begin{document}

%\title{Advection, reaction, and diffusion in a model of plankton dynamics}
%\title{Small-scale structure from advection and reaction
%in a model of plankton dynamics.\\ Tracer microstructure
%transitions in reacting fields under chaotic advection: A model of
%plankton dynamics }
%\title{Transitions in the small-scale structure of reacting fields under
%chaotic advection: Application to a model of plankton dynamics}

\title{Small-scale structure of nonlinearly interacting
species advected by chaotic flows}

\author{Emilio Hern\'{a}ndez-Garc\'{\i}a$^1$, Crist\'{o}bal
L\'{o}pez$^2$, and Zolt\'an Neufeld$^{3}$ }
\address{$^1$ Instituto Mediterr\'aneo de Estudios Avanzados (IMEDEA),
CSIC-Universitat de les Illes Balears, E-07071 Palma de Mallorca,
Spain \\ $^2$ Dipartimento di Fisica, Universit\`{a} di Roma `La
Sapienza', P.le A. Moro 2, I-00185, Roma, Italy \\ $^3$Department
of Applied Mathematics and Theoretical Physics, University of
Cambridge, Silver Street, Cambridge CB3 9EW, UK}

\maketitle

\begin{abstract}
We study the spatial patterns formed by interacting biological
populations or reacting chemicals under the influence of chaotic
flows. Multiple species and nonlinear interactions are explicitly
considered, as well as cases of smooth and nonsmooth forcing
sources. The small-scale structure can be obtained in terms of
characteristic Lyapunov exponents of the flow and of the chemical
dynamics. Different kinds of morphological transitions are
identified. Numerical results from a three-component plankton
dynamics model support the theory, and they serve also to
illustrate the influence of asymmetric couplings.
\end{abstract}
\vspace{1.5cm}

{\bf The transport of reacting substances by a fluid flow is a
problem appearing in a variety of disciplines, from classical
combustion studies to chemical reactor design. Important
environmental examples arise in the study of atmospheric advection
of reactive pollutants or chemicals, such as ozone, or in the
dynamics of plankton populations in ocean currents. The
inhomogeneous nature of the resulting spatial distributions was
recognized some time ago, but more recently satellite remote
sensing and detailed numerical simulations identify filaments,
irregular patches, sharp gradients, and other complex structures
involving a wide range of spatial scales in the concentration
patterns. We analyze here the small-scale structure of a large
class of models of transported reacting substances in terms of
basic concepts from dynamical systems theory, and apply the
results to a model of plankton dynamics. }

\vspace{1.5cm}

\section{Introduction}
\label{sec:intro}

Tracers stirred by fluid motion are known to develop strong
inhomogeneities, usually in the form of filamental features,
arising from a kind of variance cascade from the forcing scale
towards smaller scales\cite{Pierrehumbert}. These structures are
observed at size ranges as diverse as the ones relevant for
laboratory experiments\cite{sommerer,Gollub99,Gollub01},
atmospheric transport\cite{Balluch,tuck}, or temperature or
chlorophyll patchiness in the ocean\cite{patchiness,Mackas85}.
They provide sensitive mixing mechanisms and they are in some
sense ``catalists" for chemical or biological activity occurring
in the flows\cite{TelPNAS,Toroczkai}. For example, in the case of
atmospheric chemistry, the presence of strong concentration
gradients has been shown to have profound impact on global
chemical time-scales \cite{legras}. The same phenomenon has been
observed in models of ocean plankton dynamics\cite{mahadevan}. On
the other hand, chemical reactions or biological competition
occurring in the advected substances modifies the characteristics
of its spatial distribution\cite{Denman76,Abraham,Zoltan01}. Thus
the interplay between fluid flow and its chemical activity is an
important issue to understand the spatial structure of
concentration fields both in environmental and in artificial
flows.

For the case of inert transported tracers (also called the {\sl
passive scalar} problem) much progress has been achieved in the
last years in describing, at least in statistical terms, its
spatial structure \cite{Shraiman,rmpfalko}. Most of the results
have been formulated in terms of structure functions, which
describe the statistics of spatial fluctuations, or of variance
power spectra. A regime of fluid motion for which considerable
understanding has been attained is the so called {\sl Batchelor
regime}\cite{batchelor}, or {\sl viscous-convective range}, which
is the range of scales in a turbulent flow below the Kolmogorov
scale for the velocity field (the flow is thus dominated by
viscosity and spatially smooth) but above the scale for which
diffusion effects smooth out the structure of the transported
substance. This regime is of appreciable extent for large Schmidt
number (Prandtl number if the transported substance is
temperature). It is by now clear that transport behavior in this
range is equivalent to the one generically obtained under the
conditions of {\sl Lagrangian chaos} or {\sl chaotic advection}
\cite{aref,crisanti,oreste}, that is, stretching and folding of
fluid elements following chaotic trajectories in laminar flows.
Simple two-dimensional flows, and even steady flows in three
dimensions, lead generally to this kind of fluid particle motion.
In addition to modeling laboratory situations, the chaotic
advection paradigm has been shown to be a useful approach to
understand geophysical transport processes at large scales
\cite{peter}. In the Batchelor or chaotic advection regime, in
closed flows, the power spectrum presents a $k^{-1}$ decay at
large wavenumbers\cite{batchelor,vulpiani}, and the structure
functions behave logarithmically \cite{Chertkov95,tabeling}. In
open flows the singularities are restricted to fractal sets of
zero measure\cite{jung,openflows}, but they can be experimentally
visualized\cite{sommerer} and affect the global scaling behavior.

In the case of reacting transported substances (still passive in
the sense that they do not alter the flow), simple model reactions
such as $A+B
\rightarrow 2B$ \cite{Metcalfe} or $A+B\rightarrow C$ were studied
in both closed and open \cite{Toroczkai} flow situations.
Possibly, one of the simplest chemical reaction models one can
consider is the first-order reaction, or linearly decaying tracer.
It consists simply in the decrease of the concentration of a
chemical at a rate proportional to the same concentration. This
dynamics describes the consumption of a reactant in a binary
reaction, when the other reactant concentration is kept constant,
or the spontaneous decay of a finite-lifetime substance, such as a
radioactive tracer or a fluorescent dye. In the presence of a
source the concentration reaches a non-zero statistically steady
state. Sea-surface temperature relaxation towards atmospheric
values\cite{abrahamdecay}, and the relaxation of plankton and
nutrient concentrations towards mixed-layer values\cite{mahadevan}
have also been modeled in this way. Ref.~\cite{Nam2000} pointed
out the relevance of this model for vorticity dynamics in
two-dimensional turbulence in the presence of drag. Recent results
\cite{PRL,PRE,Chertkov98,Nam1999} have completed the classical
ones by Corrsin\cite{Corrsin} so that we have now a rather
complete picture of the spatial structure of finite-lifetime
substances transported by chaotic flows. The decay-time constant
makes the tracer power spectrum steeper than the Batchelor law
obtained for the passive tracer, and in general the scaling of the
structure functions is controlled by the relationship between the
decay-time constant and the Lagrangian Lyapunov exponent of the
flow. The basic physical mechanisms behind these results are the
compression of fluid elements along contracting directions of the
flow, with the consequent increase in the gradients, and the
tendency of the decaying chemistry to relax these gradients. The
competition between these processes leads to the appearance of
morphological transitions as the value of the two time scales
changes: when the chemical decay is faster than the compression by
the flow, the concentration pattern is differentiable and
characterized by a smooth structure. In the opposite case the
concentration develops a rough (non-differentiable) structure
which reflects in non trivial scaling exponents for the structure
functions. The rough distributions have filamental aspect, since
there is always an expanding direction for the flow along which
the concentration is smoothed out. Thus these transitions have
been termed {\sl smooth-filamental} transitions\cite{PRL}.

This simple picture should be completed with the recognition that
stretching (and compression) is usually not homogeneous, and
different parts of the flow experience different stretching
histories characterized by the probability distribution of the
Lagrangian finite-time Lyapunov exponents of the flow. This has as
a consequence that the set of structure functions display
anomalous (multifractal) behavior\cite{Chertkov98,PRE}. Different
points in the system may display different scaling behavior, but
the morphological transitions mentioned above can be still
identified as the change of behavior in a macroscopic portion of
the system, that is, a part with fractal dimension equal to the
full spatial dimension. The anomalous scaling is particularly
pronounced in the case of open flows with a localized mixing
region.

The presence of the above intermittency corrections to simple
scaling behavior, both in inert and in reacting flows, has a
particular status: on the one hand it implies that some of the
assumptions implicit in the simplest theoretical approaches are
incorrect at a fundamental level; on the other hand, the
corrections in quantities most directly accessible to experiments,
such as low-order structure functions, could be rather small.
Thus, it is usually a rather good approximation to neglect, in a
first approach to the problem, multifractal behavior and
concentrate into the bulk, dominant behavior. This is specially
pertinent in environmental flows, where precise quantitative
measurements are not always possible or reproducible. Comparison
between structure functions of different orders has been however
performed, and multifractality clearly demonstrated
there\cite{multifractal1,multifractal2}.

In addition to its intrinsic value as a model for several relevant
chemical situations, the linearly decaying model has been pointed
out to represent a much broader class of chemical
reactions\cite{PRL,Chertkov99}: essentially any chemistry (or
biology) leading to a negative Lagrangian chemical Lyapunov
exponent, as long as we consider just the small-scale structure.
The Lagrangian chemical Lyapunov exponent is defined as the
average rate of convergence or divergence between concentration
values present in a particular fluid particle, when it is
initialized with slightly different initial concentration values.
It turns out that it is enough to replace the decay rate of the
linearly decaying chemical model by the largest (less negative) of
the chemical Lyapunov exponents to obtain the small-scale
characterization of the structure of a nonlinear chemical or
biological model. Although some qualitative numerical checks of
this equivalence have been already presented\cite{PCE} both for
open as for closed flows, no quantitative evaluation of the
predictions of the theory has been presented so far.

In this Paper, we consider the scaling behavior of reacting fields
advected by Lagrangian chaotic flows. We consider multicomponent
nonlinear chemical or biological models, and generalize previous
results to include the possibility of non-smooth source terms. The
concept of {\sl smooth-filamental} transition is revisited in the
presence of these non-smooth forcings, which lead to different
kinds of patterns. The theory is tested with numerical simulations
of a model of oceanic plankton dynamics in a simple
two-dimensional closed flow. Part of our interest in addressing
multicomponent nonlinearly interacting populations arises from the
need to clarify differences which seem to be observed in the
scaling behavior of different plankton species stirred by the same
flow\cite{Mackas79,Steele92,Abraham}. Our results indicate that
such differences may arise, but, as far as only the small-scale
behavior and chaotic (non-turbulent) advection are considered,
they require rather asymmetrical couplings. In the theoretical
developments we address the main scaling behavior and neglect any
intermittency correction. This allows us to concentrate into the
different morphological transitions, which we believe are the most
relevant predictions to be compared with experiment or
observations. The comparison with the numerical results is
performed however in terms of the first-order structure function.
The agreement between the theory and the numerics confirms that
multifractal corrections to this low-order structure function are
weak.

The Paper is organized as follows: After this Introduction, we
present in Sect. \ref{sec:structure} the general results for the
small-scale structure of the advected fields, obtained from
consideration of the Lagrangian properties along particle
trajectories. Particular plankton and flow models are presented in
Sect. \ref{sec:numerics}, which also contains numerical results
which support the theory and illustrate the variety of transitions
arising from the interplay between reaction, advection and
forcing.  We conclude in Sect. \ref{sec:conclusions} with a
summary and open questions.

\section{The small-scale structure of advected chemicals}
\label{sec:structure}

\subsection{Evolution models and their Lagrangian representation}
\label{subsec:lagrangian}

The spatiotemporal evolution of reacting fields is determined by
advection-reaction-diffusion equations. Advection because they are
under the influence of a flow, reaction because we consider
species interacting with themselves and/or with the carrying
medium. Diffusion because turbulent or molecular random motion
smoothes out the smallest scales. For the case of an
incompressible velocity field ${\bf v}({\bf r},t)$, the standard
form of these equations is
\BE
\frac{\partial \bc(\bx,t)}{\partial t} + \bv (\bx,t)\cdot \nabla \bc(\bx,t) =
\bF(\bc,\bx)+ D \nabla ^2 \bc(\bx,t)  \ .
\label{arde}
\EE
We consider our system to be confined in a $d$-dimensional box.
The velocity field is described by the $d$-dimensional vector
field $\bv(\bx,t)$. The particular numerical example of
Sect.~\ref{sec:numerics} will be a time-dependent smooth
two-dimensional flow. $\bc(\bx,t)=(c_1(\bx,t),c_2(\bx,t), ...,
c_N(\bx,t) )$ is the set of $N$ interacting chemical or biological
fields advected by the flow, and $(F_1, ..., F_N)
\equiv \bF = \bF(c_1,...,c_N,\bx)$ are the functions accounting for the
interaction of the fields (e.g. chemical reactions or
predator-prey interactions) including sources. The explicit
dependence on the spatial coordinate $\bx$ accounts for
inhomogeneous distributions of sources and sinks of the
substances, or for the spatial dependence of reaction
coefficients, which act as an external forcing on the
concentrations. If no such forcing is present, the final
distributions will be finally homogeneized by diffusion, and
spatial pattern will occur only during a transient regime. In the
presence of forcing, some statistical steady state will be
attained from the input of substances through the sources and
sinks, and the transport of these inhomogeneities to smaller
scales by advection. There is no difficulty in considering an
explicit time-dependence in the sources: $\bF = \bF(\bc,\bx,t)$.
For simplicity in the notation we do not write explicitly this
dependence, but it is easy to see that our results (in particular
Eq.~(\ref{final}) below) are not altered by this.

The scaling properties of the concentration fields $\bc$ would
depend on the smoothness properties of both the velocity field
$\bv$ and of the interactions $\bF$. For the velocity field we
consider a smooth spatial behavior, which corresponds to the
situation of chaotic advection, or to the Batchelor regime in a
high Schmidt number turbulent flow. In previous works
\cite{PRL,PRE} we have assumed also a smooth dependence of $\bF$
both on the concentrations and in space. This allowed us to write
\BE
\bF(\bc+\delta\bc,\bx+\delta\bx)=\bF(\bc,\bx)+{\bf
DF}(\bc,\bx)\cdot\delta\bc+\nabla\bF(\bc,\bx)\cdot\delta\bx + ...
\label{Fsmooth}
\EE
The ellipsis denotes higher order terms. ${\bf DF}$ is the
(square) matrix made of the derivatives of the $N$ components of
$\bF$ with respect to the $N$ concentrations $\bc=(c_1, ...,
c_N)$. $\nabla\bF$ is the (rectangular) matrix of derivatives of
the components of $\bF$ with respect to the spatial coordinates.

There are situations in which the spatial dependence of $\bF$ is
not smooth. This will occur when source terms arise from physical
processes that lead themselves to nontrivial scaling properties.
For example, water temperature is an important variable that
controls growth rates in plankton models. But it is itself a
quantity which is transported by the flow and subjected to
processes that provide to it a nonsmooth spatial structure. Then,
source terms depending on temperature may be not well modeled by
(\ref{Fsmooth}). Other examples have been already presented in the
literature, as the case of plankton grazing on a multifractal
nutrient field \cite{Marguerit98}. In most cases the structure of
all the fields will be determined by the combined effect of their
mutual coupling and of the flow. But in cases in which couplings
are unidirectional, that is, one of the components influences the
others but there is no feedback on it, the influence of the
component which does not receive feedback is better modeled as a
source term forcing on the remaining components. An explicit
example of this situation will be presented in Sect.
\ref{sec:numerics}.

For the case of nonsmooth source forcing, we will use
\BE
\bF(\bc+\delta\bc,\bx+\delta\bx)=\bF(\bc,\bx)+{\bf
DF}(\bc,\bx)\cdot\delta\bc+\delta_xF + ...
\label{Fnonsmooth}\EE
instead of (\ref{Fsmooth}). $\delta_xF=\delta_xF(\bc,\bx,\bdx)$ is
the set of spatial increments $\delta_xF=(\delta_xF_1, ...,
\delta_xF_N)$, with $\delta_xF_i \equiv
F_i(\bc,\bx+\delta\bx)-F_i(\bc,\bx)$, which scale at small
distances as
\BE
|\delta_xF_i| \sim |\bdx|^{\beta_i} \ .
\label{scalingdF}
\EE
$\beta_i=1$ for smooth sources. In cases in which $\bF$ is bounded
as a function of space, as in the concrete model to be discussed
below, the behavior (\ref{scalingdF}) should cross over a
saturation behavior for increasing $|\bdx|$. We call $L$ the
length scale at which this will happen (we assume it to be the
same for all chemical sources labeled by different values of the
index $i$). It should be of the order of the largest scale of
inhomogeneity in the source terms $\delta_x\bF$, usually of the
order of the system size. The saturation value of $|\delta_xF_i|$
will be then of the order of $L^{\beta_i}$.

 Diffusion effects are only important at the smallest
scales and we will neglect them in the following. In this limit of
zero diffusion $D
\to 0$
 the above description can be
recast in Lagrangian form. To this end we introduce some notation:
$\bX_{\tinit}^t(\bx_0)$ will denote the position at time $t$ of a
fluid particle that at some initial time $\tinit$ was at $\bx_0$,
that is, it is the trajectory solution of
\BE
\frac{d}{dt}\bX_{\tinit}^t=\bv(\bX_{\tinit}^t,t)
\label{trajectories}
\EE
satisfying $\bX_{\tinit}^{\tinit}(\bx_0)=\bx_0$. The set of
Lagrangian concentrations inside this fluid element will be
denoted by $\bC_{\tinit}^t(\bx_0)$. The relationships between the
Eulerian and Lagrangian concentrations are
\BE
\bC_{\tinit}^t(\bx_0) = \bc\left(\bx=\bX_{\tinit}^t(\bx_0),t\right)
\EE
\BE
\bc(\bx,t)=\bC_{\tinit}^t\left(\bx_0=\bX_t^{\tinit}(\bx)\right)  \
.
\EE
The last expression states that the relevant particle trajectories
to recover Eulerian concentrations at position $\bx$ are the ones
that end at $\bx$ at time $t>\tinit$, that is, solutions of
Eq.~(\ref{trajectories}) integrated backwards in time from {\sl
final} conditions at $\bx$.

Equation (\ref{arde}) in the Lagrangian framework, that is the
equation ruling the chemical dynamics inside the fluid element
which follows the trajectory $\bX_{\tinit}^t(\bx_0)$, reads
\begin{equation}
\frac{d}{dt}\bC_{\tinit}^t(\bx_0)=\bF\left( \bC_{\tinit}^t(\bx_0),
\bx=\bX_{\tinit}^t(\bx_0)
\right)   \ .
\label{chemical}
\end{equation}

The set of equations (\ref{trajectories}) and (\ref{chemical}),
which we call the {\sl flow} and the {\sl chemical} subsystem,
respectively, are the basic starting point for our analysis. Note
that the coupling between both equations only appears if there is
space dependence in $\bF$, that is, if there are inhomogeneous
sources.

The quantity we are interested in is the difference in
concentration between neighboring points:
\BE
\delta \bc(\bx,t;\delta\bx) \equiv \bc(\bx+\bdx,t)-\bc(\bx,t)
\label{deltac}
\EE
and in particular in its scaling behavior at small distances (but
larger than the diffusion scale, which we are neglecting), which
defines the set of H\"older exponents $\alpha_i$:
\BE
|\delta c_i(\bx,t;\delta\bx)| \sim |\bdx|^{\alpha_i}
\label{Holder}
\EE
We will consider here only the large-time statistically steady
state obtained under forcing. Thus the time in (\ref{Holder})
should be considered to be large enough  for the initial
concentrations at $t_0$ to be forgotten. Only after this
$t\rightarrow \infty$ limit should the small-$|\bdx|$ behavior be
considered.

We allow for different scaling behavior in each concentration
$c_i$, although we will see that this will not be the usual case.
In general, there will be space- and time-dependent prefactors in
(\ref{Holder}), but one expects the exponent, as any other
statistical characteristic of the concentration pattern, to be a
constant in space and time. Under the hypothesis of our approach,
this is indeed the case. It should be said, however, that because
of the intermittency corrections we will neglect, $\alpha_i$ may
have values different from the constant calculated below in sets
of points of zero measure.

The Lagrangian quantity analogous to (\ref{deltac}) is the
difference in concentration between two fluid particles:
\BE
\delta\bC_{\tinit}^t(\bx_0;\bdx_0) \equiv
\bC_{\tinit}^t(\bx_0+\bdx_0)-\bC_{\tinit}^t(\bx_0) \ .
\label{deltaClag}
\EE

\subsection{Lyapunov characterization of the flow subsystem}
\label{subsec:flow}

We introduce the difference between two trajectories that start at
points separated by $\bdx_0$:
\BE
\delta\bX_{\tinit}^t(\bx_0;\bdx_0) \equiv
\bX_{\tinit}^t(\bx_0+\bdx_0)-\bX_{\tinit}^t(\bx_0)  \ .
\label{deltaX}
\EE
If this quantity is small, its dynamics is given by the
linearization of Eq.~(\ref{trajectories}), that is
\BE
\frac{d}{dt}\delta\bX_{\tinit}^t=
{\bf J}(\bX_{\tinit}^t)\cdot \delta\bX_{\tinit}^t  \ ,
\label{Xlin}
\EE
where ${\bf J}$ is the Jacobian matrix of the derivatives of $\bv$
with respect to the spatial coordinates. The solution of
(\ref{Xlin}) can be written as
\BE
\delta\bX_{\tinit}^{t_2}(\bx_0;\delta\bx_0)={\bf T}(t_2,t_1) \cdot
\delta\bX_{\tinit}^{t_1}(\bx_0;\delta\bx_0)
\label{dXsol}
\EE
in terms of the $d \times d$ fundamental matrix ${\bf T}$, which
is the matrix solution of
\BE
\frac{d}{dt_2}{\bf T}(t_2,t_1)=
{\bf J}(\bX_{\tinit}^{t_2})\cdot {\bf T}(t_2,t_1)
\label{fundamentalX}
\EE
with initial condition ${\bf T}(t_1,t_1)={\bf I}$, the $d \times
d$ identity matrix.

Equation (\ref{Xlin}) is the well-known variational equation
associated to the flow (\ref{trajectories}). Some hypothesis about
the flow should be made to obtain concrete information. A
convenient assumption is to assume that (\ref{trajectories}) is a
hyperbolic and ergodic dynamical system \cite{EckmannRMP}. This
means that at every point of the system one can identify
contracting and expanding directions, associated with Lyapunov
exponents $\lambda_1>\lambda_2> ...>\lambda_d$, which give
exponential growth or decay of $|\delta\bX_{\tinit}^t|$ at long
times. Common flows are never perfectly hyperbolic, and there are
points at which, because of the presence of KAM tori, or because
of tangencies between stable and unstable manifolds, the
characteristic directions are not defined. For {\sl sufficiently
chaotic} flows such situations are not frequent, and it is a good
approximation to assume full hyperbolicity. This approximation
allows considerable generality in the analysis, and will be
implicit in the following. Since the flow is incompressible, the
sum of the Lyapunov exponents is zero, and thus for chaotic flows,
$\lambda_1
> 0$ and $\lambda_d<0$. For two-dimensional flows,
$\lambda_2=-|\lambda_1|$. For $t_2 \gg t_1$, the singular value
decomposition of the matrix ${\bf T}(t_2,t_1)$ will be dominated
by the largest eigenvalues, which are related to the Lyapunov
exponents, so that the action of ${\bf T}$ on a generic
displacement vector $\bdx$ will be given by
\BE
{\bf T}(t_2,t_1)\cdot \bdx \approx \bpi_1(t_2)e^{\lambda_1
(t_2-t_1)}\bpi_1^\dagger(t_1)\cdot \bdx   \ .
\label{Tlongt+}
\EE
$\bpi_1$ and $\bpi_1^\dagger$ are the right and left eigenvectors,
respectively, associated to the singular value $e^{\lambda_1
(t_2-t_1)}$. In physical terms, they are the unstable directions
attached to the points $\bx=\bX_{\tinit}^{t_2}$ and
$\bx=\bX_{\tinit}^{t_1}$, respectively. Analogously, if $t_2
\ll t_1$, the singular value decomposition of ${\bf T}(t_2,t_1)$
will be dominated by the most negative Lyapunov exponent
$\lambda_d$, and:
\BE
{\bf T}(t_2,t_1)\cdot \bdx \approx \bpi_d(t_2)e^{\lambda_d
(t_2-t_1)}\bpi_d^\dagger(t_1)\cdot \bdx    \ .
\label{Tlongt-}
\EE
We note, and this will be relevant in our results, that at each
time $t_1$, Eqs. (\ref{Tlongt+}) and (\ref{Tlongt-}) will be valid
for all orientations of $\bdx$ {\sl except for orientations
perpendicular to $\bpi_d^\dagger(t_1)$}. Along these directions,
the action of {\bf T} will be associated to subdominant Lyapunov
exponents, as discussed below.

If $|t_2-t_1| \rightarrow \infty$, expressions (\ref{Tlongt+}) or
(\ref{Tlongt-}) would indicate that $\delta\bX_{\tinit}^t$ grows
without limit. At some moment the linearized evolution
(\ref{Xlin}) will not longer be valid, and $\delta\bX_{\tinit}^t$
will saturate at a value of the order of system size, or of some
characteristic length scale of the velocity field. We call this
length scale $L$. For simplicity we use the same symbol for it as
for the length scale of saturation of $\delta_x\bF$. If they are
different, the discussion after Eq.~(\ref{deltaci}) below implies
that only the smallest of these length scales enters into the
analysis. Saturation of $\delta\bX_{\tinit}^t$ will happen at a
time $\tau=\tau(\delta\bx_0)$ such that
$\delta\bX_{\tinit}^\tau(\bx_0;\bdx_0) \approx L$, or
\BE
\tau(\bx_0) \approx -\frac{1}{\lambda}\log\frac{|\bdx|}{L}  \ ,
\label{tau}
\EE
where $\lambda$ is either $\lambda_1>0$ if we are using
(\ref{Tlongt+}) so that $\tau>0$, or $\lambda_d<0$, if we are
looking for the evolution towards the past (\ref{Tlongt-}) so that
$\tau<0$.

It should be noted that both (\ref{Tlongt+}) and (\ref{Tlongt-})
give only the typical asymptotic behavior at large time
differences. It needs to be corrected at least in two aspects,
even within the hyperbolicity hypothesis implicit in our approach:
On the one hand, at finite times the Lyapunov exponent has not
reached completely its long-time value, but it has a value which
depends on the initial condition and has a characteristic
probability distribution \cite{Ott,VulpiBook}. On the other hand,
even at infinite times, there is a (fractal) set of spatial points
for which the Lyapunov exponent may differ from the typical
asymptotic value. This set has zero measure because the
distribution of finite-time Lyapunov exponents becomes narrower
and narrower in time, but its presence may affect some of the
scaling behaviors described below. These two features are not
independent. Both arise from the characteristic slow approach of
the Lyapunov exponent towards its asymptotic long-time
behavior\cite{Goldhirsch87}, and their consequences are also the
same: the introduction of intermittency corrections to the scaling
behavior. Following our goal of concentrating just in bulk scaling
and transition behavior, we do not consider in the following any
correction to (\ref{Tlongt+}) or (\ref{Tlongt-}).

\subsection{Scaling behavior of the chemical subsystem}
\label{subsec:chemical}

From (\ref{chemical}), the concentration difference
(\ref{deltaClag}) satisfies
\BA
&\frac{d}{dt}&\delta\bC_{\tinit}^t(\bx_0;\bdx_0) = \nonumber \\
&&\bF\left(\bC_{\tinit}^t(\bx_0+\bdx_0),
\bx=\bX_{\tinit}^t(\bx_0+\bdx_0)\right) - \bF\left(\bC_{\tinit}^t(\bx_0),
\bx=\bX_{\tinit}^t(\bx_0)\right) =   \nonumber  \\
&&\bF\left(\bC_{\tinit}^t(\bx_0)+\delta\bC_{\tinit}^t(\bx_0;\bdx_0),
\bX_{\tinit}^t(\bx_0)+\delta\bX_{\tinit}^t(\bx_0;\bdx_0)\right) -
\bF\left(\bC_{\tinit}^t(\bx_0),
\bX_{\tinit}^t(\bx_0)\right)    \ .
\label{ddeltaCdt}
\EA
For all times during which $\delta\bC_{\tinit}^t$ and
$\delta\bX_{\tinit}^t$ remain sufficiently small, we can use
(\ref{Fnonsmooth}) to get
\BA
&\frac{d}{dt}&\delta\bC_{\tinit}^t(\bx_0;\bdx_0)=   \nonumber \\
&{\bf DF}&\left(   \bC_{\tinit}^t(\bx_0)  ,  \bX_{\tinit}^t(\bx_0)
\right) \cdot
\delta\bC_{\tinit}^t(\bx_0;\bdx_0) + \delta_x\bF\left(
\bC_{\tinit}^t(\bx_0),\bX_{\tinit}^t(\bx_0);\delta\bX_{\tinit}^t(\bx_0;\bdx_0)
\right)   \ .
\label{ddeltaCdtlin}
\EA
This is a {\sl linear} equation for $\delta\bC_{\tinit}^t$, even
though the complete dynamics (\ref{chemical}) may be nonlinear.

The general solution of this linear system may be written in terms
of its fundamental matrix ${\bf M}(t_2,t_1)$, which is the $N
\times N$-matrix solution of
\BE
\frac{d}{dt_2}{\bf M}(t_2,t_1)
= {\bf DF}\left( \bC_{\tinit}^{t_2},
\bX_{\tinit}^{t_2}\right)\cdot {\bf M}(t_2,t_1)
\label{fundamentalC}
\EE
with initial condition ${\bf M}(t_1,t_1)=\bf I$, the identity
matrix. As before, the homogeneous part of the linearization
(\ref{ddeltaCdtlin}), or (\ref{fundamentalC}), is the variational
equation associated to the chemical subsystem (\ref{chemical}). It
defines a set of Lyapunov exponents, which we call the {\sl
chemical} Lyapunov exponents. They describe the sensitivity to
concentration initial conditions under a fixed trajectory
$\bX_{\tinit}^t(\bx_0)$. We will consider here the situation in
which all of them are negative. Positive chemical Lyapunov
exponents lead to strong divergences at small scales and in such
situation neglecting diffusion may not be justified. A treatment
related to the present one but for map models which may have
positive Lyapunov exponents can be found in \cite{LogisticChaos}.
If $t_2
\gg t_1$, the dominant term in the singular value decomposition of
{\bf M} is related to the largest (less negative) chemical
Lyapunov exponent that we denote by $\lambda_C$. The action of
this matrix on a generic vector $\delta\bc$ of the tangent space
of concentration increments would be:
\BE
 {\bf M}(t_2,t_1)\cdot \delta\bc \approx
\bmu(t_2)e^{\lambda_C(t_2-t_1)}\bmu(t_1)^\dagger \cdot \delta\bc
\ .
\label{Mlongt}
\EE
$\bmu$ and $\bmu^\dagger$ are the right and left eigenvectors of
${\bf M}(t_2,t_1)$ associated to the eigenvalue
$e^{\lambda_C(t_2-t_1)}$. As before, we neglect any fluctuation in
the value of the chemical Lyapunov exponent.

The quantity we are interested in is not really (\ref{deltaClag}),
but the Eulerian increments (\ref{deltac}). The later can be
obtained from the former:
\BE
\delta\bc(\bx,t;\bdx)=\delta\bC_{\tinit}^t\left(   \bX_t^{\tinit}(\bx);
\delta\bX_t^{\tinit}(\bx;\bdx)\right)  \ .
\label{dcdC}
\EE
From (\ref{ddeltaCdtlin}), (\ref{fundamentalC}) and (\ref{dcdC}):
\BE
\delta\bc(\bx,t;\bdx)=
{\bf M}(t,0)\cdot \delta\bc_0(\bX_t^0(\bx);\delta\bX_t^0(\bx))+
\int_0^t ds {\bf M}(t,s)\cdot
\delta_x\bF\left(\bC_t^s(\bx),\bX_t^s(\bx);
\delta\bX_t^s(\bx;\bdx)\right) \ .
\label{deltacsol}
\EE
Here and in the following we use $t_0=0$. $\delta\bc_0$ is the
concentration difference at $t=0$. The first term describes the
evolution of the initial concentration difference under the
autonomous part of the  chemical dynamics, whereas the integral
term describes the cumulative effect of the source forcing. Since
we address the long-time statistically steady state, and given
that ${\bf M}$ behaves as (\ref{Mlongt}) with negative
$\lambda_C$, the initial condition will be forgotten and we need
only to consider the integral term. In the absence of forcing, the
first term in (\ref{deltacsol}) would give rise to chemically
decaying analogs of the `strange eigenmodes' of
\cite{Pierrehumbert,Gollub99}. At large $t$, the $i$-component of
(\ref{deltacsol}) with (\ref{Mlongt}) reads
\BE
\delta c_i(\bx,t;\bdx) \approx
\int_0^t ds \left(\bmu(t)\right)_i e^{\lambda_C (t-s)} \bmu^\dagger(s)\cdot
\delta_x\bF\left(\bC_t^s(\bx),\bX_t^s(\bx);
\delta\bX_t^s(\bx;\bdx)\right)  \ .
\label{deltaci}
\EE
$\left(\bmu(t)\right)_i$ is the $i$-component of the vector
$\bmu(t)$, i.e., the component associated to the chemical species
$c_i$. Generically, the scaling behavior of the increment
(\ref{deltaci}) of the chemical species $c_i$, for any $i$ at
small $\bdx$, will be dominated by the most important component of
$\delta_x\bF$ (the one with {\sl minimum} scaling exponent,
$\beta_{m}$). This behavior will be calculated in the following.
There are however situations in which this would not be the
relevant behavior: if the $i$-component of the vector $\bmu(t)$ is
vanishing, or if the most important component of $\delta_x\bF$ at
small $\bdx$ has no projection on $\bmu^\dagger$, then subdominant
terms and Lyapunov exponents different from $\lambda_C$ should be
taken into consideration. Since the eigenvectors $\bmu$ and
$\bmu^\dagger$ are changing in time, this singular situation will
not occur generically {\sl unless} some special form of the
couplings in the model enforce this situation at all times. The
particular model to be discussed in Sect. \ref{sec:numerics} has
this property. In this Section, we analyze just the generic
behavior. We split the integral into two contributions, one during
which the backwards trajectory $\delta\bX_t^s$, and thus
$\delta_x\bF$, is growing, which corresponds to the time interval
$(t-|\tau(\bdx)|,t)$, and the rest of the time
$(0,t-|\tau(\bdx)|)$, during which the value of $\delta_x\bF$ is
saturated.  Since we are using backwards trajectories, the most
negative Lyapunov exponent $\lambda_d$ should be used in the
expression (\ref{tau}) for $\tau$. We now substitute
(\ref{Fnonsmooth}) in Eq.~(\ref{deltaci}), and then insert the
asymptotic behavior of $\delta\bX_t^s(\bx;\delta\bx)={\bf
T}(s,t)\cdot
\delta\bx$ for $t \gg s$ (Eq.~(\ref{Tlongt-})). The result is
\BE
|\delta c_i| \sim L^{\beta_m}
\int_0^{t-|\tau(\bdx)|} e^{-|\lambda_C|(t-s)} ds +
|\delta\bx|^{\beta_{m}}
\int_{t-|\tau(\bdx)|}^t
e^{\left(|\lambda_d|\beta_{m}-|\lambda_C|\right) (t-s)} ds \ .
\label{generic}
\EE
We have omitted a number of space- and time-dependent factors.
They give space and time dependence to the prefactors present in
$\delta c_i$, but they do not affect the scaling behavior. To
analyze the small scale behavior, we first take $t\rightarrow
\infty$ (changing variables to $u\equiv t-s$ is useful for this)
and then analyze the scaling behavior of each term for small
$|\delta\bx|$: The first integral always behaves as
$|\delta\bx|^{|\lambda_C|/|\lambda_d|}$ whereas the second one has
also this behavior if $|\lambda_C|<|\lambda_d|\beta_m$, and
$|\bdx|^{\beta_m}$ otherwise. The final result for the H\"older
exponents (\ref{Holder}) is then
\BE
\alpha_i=\min(\beta_m,
|\lambda_C|/|\lambda_d|)
\label{final}
\EE
We recall here that $\beta_m=\min(\beta_1,\beta_2, ...,
\beta_N)$. For differentiable sources ($\beta_m=1$) we recover the
previous result \cite{PRL,PRE}. It is also the result for a
linearly decaying chemical, if $|\lambda_C|$ plays the role of the
decay rate. An explicit time-dependence in the source terms would
affect the value of $\lambda_C$, but expression (\ref{final})
remains valid.

As stated before, the scaling exponent (\ref{final}) will be the
one obtained for generic orientation of the displacement $\bdx$.
But at each point $\bx$ there will be orientations for which the
subdominant Lyapunov exponent $\lambda_{d-1}$ should be used
instead of $\lambda_d$: the directions orthogonal to the local
most contracting direction $\bpi_d^\dagger$. Again, in directions
perpendicular to both $\bpi_d^\dagger$ and $\bpi_{d-1}^\dagger$,
$\lambda_{d-2}$ should be used, and so on. It is enough to replace
$|\lambda_d|$ by $-\lambda$ in (\ref{final}), where $\lambda$ is
the relevant Lyapunov exponent, to get the scaling behavior along
these directions. Along the expanding directions, i.e. the
directions for which the pertinent Lyapunov exponent is positive,
we have $\alpha_i=\beta_m$, which for smooth sources mean smooth
scaling behavior. We note however that source terms having
particular anisotropic properties require special consideration.
This is the case of the model in Sect.~\ref{sec:numerics}.

In the same way, at each point there are particular directions
$\delta\bc$ in concentration space such that a subdominant
chemical Lyapunov exponent should be used instead of $\lambda_C$.
But these directions will generically not be aligned with the
concentration coordinate axes, and thus would not be relevant to
the scaling of the real chemical species $\delta c_i$, unless some
particular form of the coupling is present. For the generic case,
(\ref{final}) states that the small-scale H\"{o}lder exponent of
all the interacting chemical species are the same.

The minimum condition in the equations for the H\"{o}lder
exponents give rise to interesting morphological transitions as
the model parameters are varied, as one or the other of the
expressions under the minimum function become the relevant one.
Examples of the transitions are given in the next Section.

\section{Numerical results for a plankton model}
\label{sec:numerics}

In the numerical investigations below we will consider a simple
model of plankton dynamics stirred by a two-dimensional
time-dependent flow.

%%%%%%%%%%%%%%%%%%%%%%%%%%%%%%%%%%%%%%%%%%%%%%%%%%%%%%%%%%%%%%%%%%%%%%%%
This plankton model is a typical predator-prey system
\cite{murray} where three trophic levels are considered: the
nutrients, parametrized by the carrying capacity $C$ of the water
parcel (defined as the maximum phytoplankton content it can
support in the absence of grazing), the phytoplankton $P$ and the
zooplankton $Z$ biomass concentrations. The dynamics of these
species is given by
\begin{eqnarray}
\frac{dC}{dt}&=& F_C=a \left(   C_0(\bx)-C \right)
\label{Ct}\\
\frac{dP}{dt}&=& F_P=P\left( 1-\frac{P}{C}\right)-PZ
\label{Pt}\\
\frac{dZ}{dt}&=& F_Z=PZ-b Z^2
\label{Zt}\ .
\end{eqnarray}
The Lagrangian `chemical' subsystem (\ref{chemical}) is simply
obtained by considering that all the reactions (\ref{Ct}-\ref{Zt})
occur in a particular fluid element, so that the spatial
coordinate $\bx$ in (\ref{Ct}) is taken to be the position
$\bX_{\tinit}^t$ of this fluid particle in the ocean flow. All
terms have been adimensionalized to keep a minimal number of
parameters. The first equation (\ref{Ct}) states that the carrying
capacity adapts (at a rate $a$) to the local value of some source
of nutrients $C_0$. This will be the only explicitly inhomogeneous
term in the model, and describes a spatially dependent nutrient
input arising from some oceanic topography-determined upwelling
distribution or latitude dependent illumination, for example. The
first terms in Eq.~(\ref{Pt}) describe phytoplankton logistic
growth, whereas the last one models predation (grazing) by
zooplankton. This effect gives also rise to the first term in
(\ref{Zt}). The term containing $b$, the zooplankton mortality,
describes zooplankton death due to higher trophic levels. For the
parameter values we are using, in the absence of stirring by the
flow, the system evolves to a stable equilibrium state which is
non-uniform in space because of the inhomogeneous source
$C_0(\bx)$. We use for the source the expression
$C_0(x,y)=1.7+0.52 \sin{(2
\pi x)}\cos{(2
\pi y)}$. This is a smooth function and then $\beta_C=1$.

This particular plankton dynamics has not been chosen because of
some particular biological relevance, but because it allows us to
illustrate a rather rich variety of transitions in the context of
the theory of Sect.~\ref{sec:structure}. In addition, it contains
some of the {\sl nongeneric} features that are difficult to
discuss in general, but that can be easily addressed in each
particular case by simple extensions of the theory of Sect.
\ref{sec:structure}, and that may be relevant for the
understanding of real flows. We note that the dynamics of the
carrying capacity Eq.~(\ref{Ct}) influences other species, but no
feedback from $P$ or $Z$ affects the dynamics of $C$. This leads
to a linearized dynamics described by a matrix ${\bf DF}$ which
has a box structure for all times, and thus, there is always a
Lyapunov exponent, of value exactly $-a$, associated to a
contracting direction along the direction of the carrying capacity
$\bmu=\bmu^\dagger=(1,0,0)$, which is decoupled from the ones
associated to the $P$-$Z$ remaining subsystem. This was one of the
{\sl nongeneric} situations discussed in Section
\ref{subsec:chemical}. We see that it can arise rather easily in
the presence of sufficiently asymmetric couplings between the
variables.

One way to analyze our model is to consider first the dynamics of
the carrying capacity. It is independent of the other variables
and its small-scale structure, characterized by the H\"{o}lder
exponent $\alpha_C$, will be given by the interplay between the
flow Lyapunov exponent, the decay rate $a$, and the source
smoothness degree $\beta_C=1$, with the result (\ref{final}):
\BE
\alpha_C=\min \left(  1 , \frac{a}{|\lambda_d|}
\right)  \ .
\label{alphaC}
\EE
The influence of $C$ into the remaining $P$-$Z$ subsystem appears
only through the denominator in (\ref{Pt}). We can consider the
$P$-$Z$ subsystem as a $2 \times 2$ chemical dynamics forced by a
source term $C(\bx,t)$. The associated components of $\delta_x\bF$
behave as $|\delta_xF_P|\approx |\delta C| \approx
|\bdx|^{\alpha_C}$ and $|\delta_xF_Z|=0$. We thus have that the
smoothness exponent of the forcing into this subsystem is
$\beta_m^{PZ}=\alpha_C$. From (\ref{final}), and since the
variables $P$ and $Z$ are coupled in a generic way, the H\"{o}lder
exponents describing the small-scale structure of $P$ and $Z$ are
equal and given by
\BE
\alpha_{PZ}=\min \left( \alpha_C , \frac{|\lambda_C|}{|\lambda_d|}
\right)  \ .
\label{alphaPZ}
\EE
$\lambda_C$ is now the largest Lyapunov exponent associated to the
forced $P$-$Z$ subsystem. As $\lambda_d$, it can only be estimated
numerically.

An alternative way to analyze our model is to recognize that the
theoretical arguments of Sect.~\ref{sec:structure} imply that
(\ref{final}) is valid if the minimum condition for every $i$ is
taken over the set of Lyapunov exponents which affect to this
particular species $c_i$. Considering thus the three-component
system globally forced by the smooth source $C_0$, the same
results are obtained.

We define our system to be the unit square with periodic boundary
conditions, we use the following model flow:
\begin{eqnarray}
v_x(x,y,t)& =& -\frac{2U}{T} \Theta \left(\frac{T}{2}-t \bmod T
\right) \cos({2\pi y}) \nonumber \\ v_y(x,y,t)&=& -\frac{2U}{T}
\Theta \left(t \bmod T-\frac{T}{2} \right) \cos({2\pi x})  \ .
\label{flow}
\end{eqnarray}
$\Theta(x)$ is the Heavyside step function. Note that the largest
lengths of both the source and the flow are given by the system
size, $L=1$. The flow is time-periodic with period $T$. In our
simulations we use $U=1$, which produces a single connected
chaotic region in the advection dynamics. Full hyperbolicity is
not garanteed, but the flow is chaotic enough to apply the theory
of Sect. \ref{sec:structure}. It is easy to show that the Lyapunov
exponent is inversely proportional to $T$. Since we are in two
dimensions, $\lambda_1=|\lambda_2|$. The numerically determined
value is $\lambda_1=|\lambda_2|
\approx 2.35/T$. We also fix $b=0.05$, and vary the values of $T$
and $a$.

A quantity that is usually considered in scaling studies of the
properties of fluid patterns is the $n$-order structure function.
It is defined as
\BE
S_q^{i}(\delta x) \equiv \left<   |\delta c_i(\bx,t;\bdx)|^q
\right> \ ,
\label{structurefunction}
\EE
were the average is over space. In our numerical simulations, we
perform the average in (\ref{structurefunction}) over the points
in 50 rectilinear segments across the system, or `transects'.

In general, the behavior of $S_q^{i}$ for small $\delta x$ would
be of the form
\BE S_q^{i}(\delta x) \sim |\delta x|^{\xi_q^i}  \ ,
\label{anomalousscaling}
\EE
If all the points of the system have the same value of the
H\"{o}lder exponent for the chemical species $i$, then it is clear
that $\xi_q^i = q \alpha_i$. In general, intermittency will
introduce corrections so that $\xi_q^i$ will be a nonlinear
function of $q$. For the structure functions of lower order these
corrections are usually small \cite{PRE}, so that for example for
the first order structure function we expect
\BE
S_1^{i}(\delta x) \sim |\delta x|^{\alpha_i}  \
\label{normalscaling}
\EE
to be a good approximation, where the $\alpha_i$ are given by
(\ref{alphaC}) and (\ref{alphaPZ}).

In order to compare theory and numerics, one needs to calculate
the chemical Lyapunov exponent associated to the subsystem
$P$-$Z$. This is done by integrating, using the same Lagrangian
trajectory $\bX_0^t$, two copies of (\ref{Ct})-(\ref{Zt}) with
slightly different initial values of $P$ and $Z$, but the same
initial $C$. This leads to a time dependent difference between the
two copies, $\delta P$ and $\delta Z$, from which we monitor
$(\delta P^2+\delta Z^2)^{1/2}$. This is fitted to an exponential
that grows or decays with a characteristic time which identifies
$\lambda_C$. In general, $\lambda_C$ may have an arbitrary
dependence on the model parameters. It can even change sign by
changing the characteristics of the flow an thus of the trajectory
$\delta\bX_0^t$ \cite{VulpiBook,LogisticChaos}. But in our
numerical study, for the particular flow and chemical subsystems,
fixed values of $U$ and $b$, and for all the considered range of
$T$ and $a$ to be discussed below, the value of $\lambda_C$ was
contained in the narrow interval $(-0.0386,-0.0380)$. Thus
expressions (\ref{alphaC}) and (\ref{alphaPZ}) become rather
simple functions of $a$ and $T$, since $\lambda_C$ is nearly
constant, and $|\lambda_2|$ is inversely proportional to $T$.

For $T$ large enough (slow flow), $|\lambda_2|$ is small and we
have that $\alpha_C=\alpha_{PZ}=1$: the three concentration fields
are smooth. By decreasing $T$, that is, by increasing the
chaoticity of the flow, Eqs. (\ref{alphaC}) and (\ref{alphaPZ})
predict two different sequences of transitions depending on the
relationship between $a$ and $\lambda_C$, which we now describe.

\begin{figure}
\epsfig{file=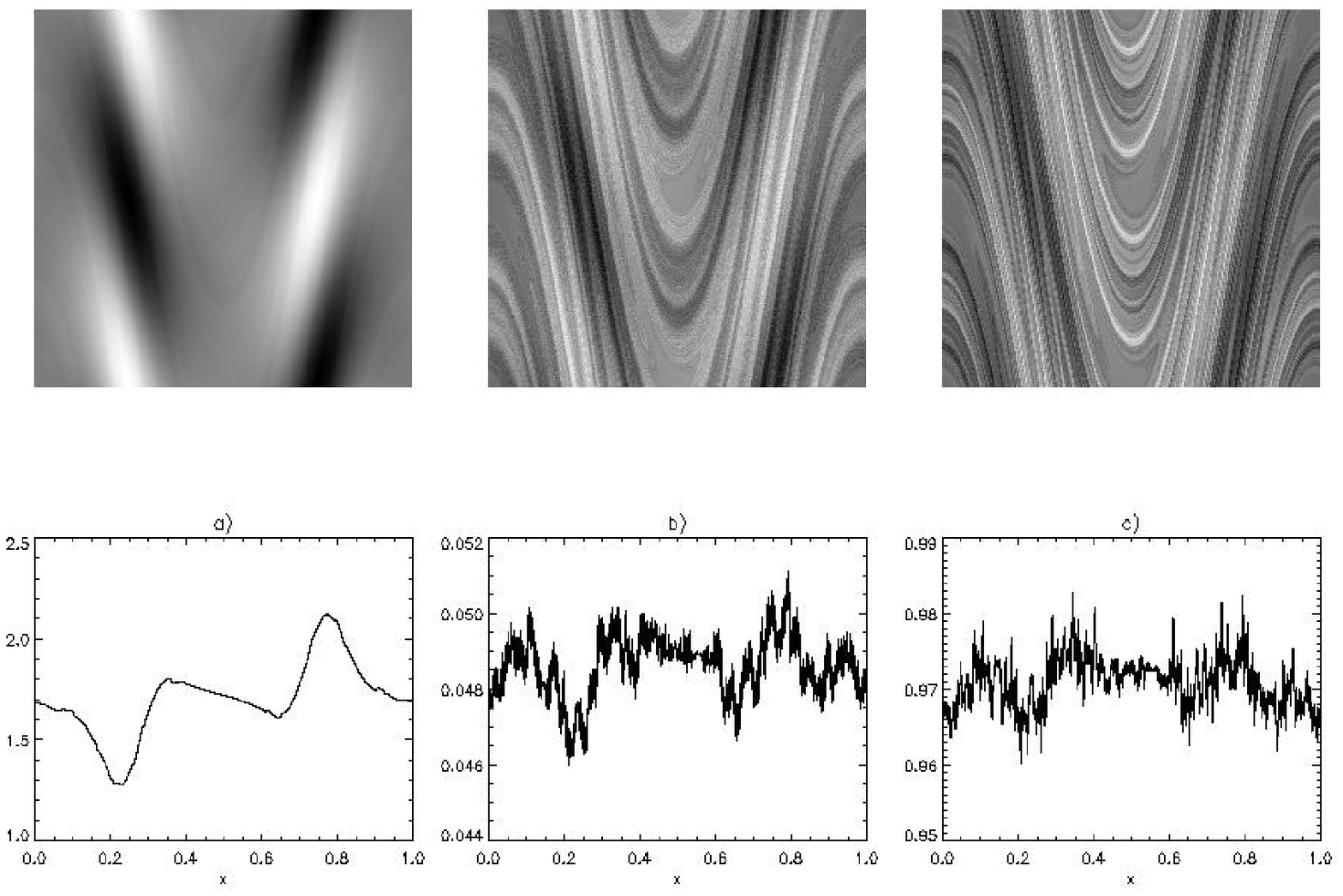,width=0.9\linewidth }
\vspace*{1 cm}
\caption{Carrying capacity (a), phytoplankton (b) and zooplankton (c) configurations
for $a=0.2$ and $T=0$. The upper panels show snapshots of the
two-dimensional system, whereas the lower ones display the
corresponding concentrations along a horizontal transect taken
along $y=0.6$. The smooth character of $C$, and the singular one
of $P$ and $Z$, is clearly seen.}
\label{fig:3patsA02T20}
\end{figure}

First, if $a$ remains larger than $|\lambda_C|$, the first
transition encountered by decreasing $T$ is the situation in which
$P$ and $Z$ become filamental, whereas $C$ remains still smooth.
We show in Figure \ref{fig:3patsA02T20} instantaneous
configurations of the three concentration fields, for $a=0.2$,
$T=20$, that is, after crossing that transition\cite{numerics}.
The predicted smooth character of $C$ and filamental of $P$ and
$Z$ is clearly observed. As expected, the features become aligned
with the stable and unstable manifolds of the flow. These
manifolds change periodically in time, following the periodicity
of the flow, but the scaling properties of them do not change.
There are differences in the details of the structure of $P$ and
$Z$, but Fig.~\ref{fig:structA02T20}, in which we show the
first-order structure functions, shows that the small scale
behavior is similar. The theoretical predictions for the measured
value of $|\lambda_C|\approx 0.0386$, $\alpha_C=1$ and
$\alpha_{PZ}\approx 0.33$, are also shown, in very good agreement
with the numerics (least-squares fitting gives $\alpha_C\approx
1.0$ and $\alpha_{PZ}\approx 0.35$), despite the approximations
involved.

\begin{figure}
\epsfig{file=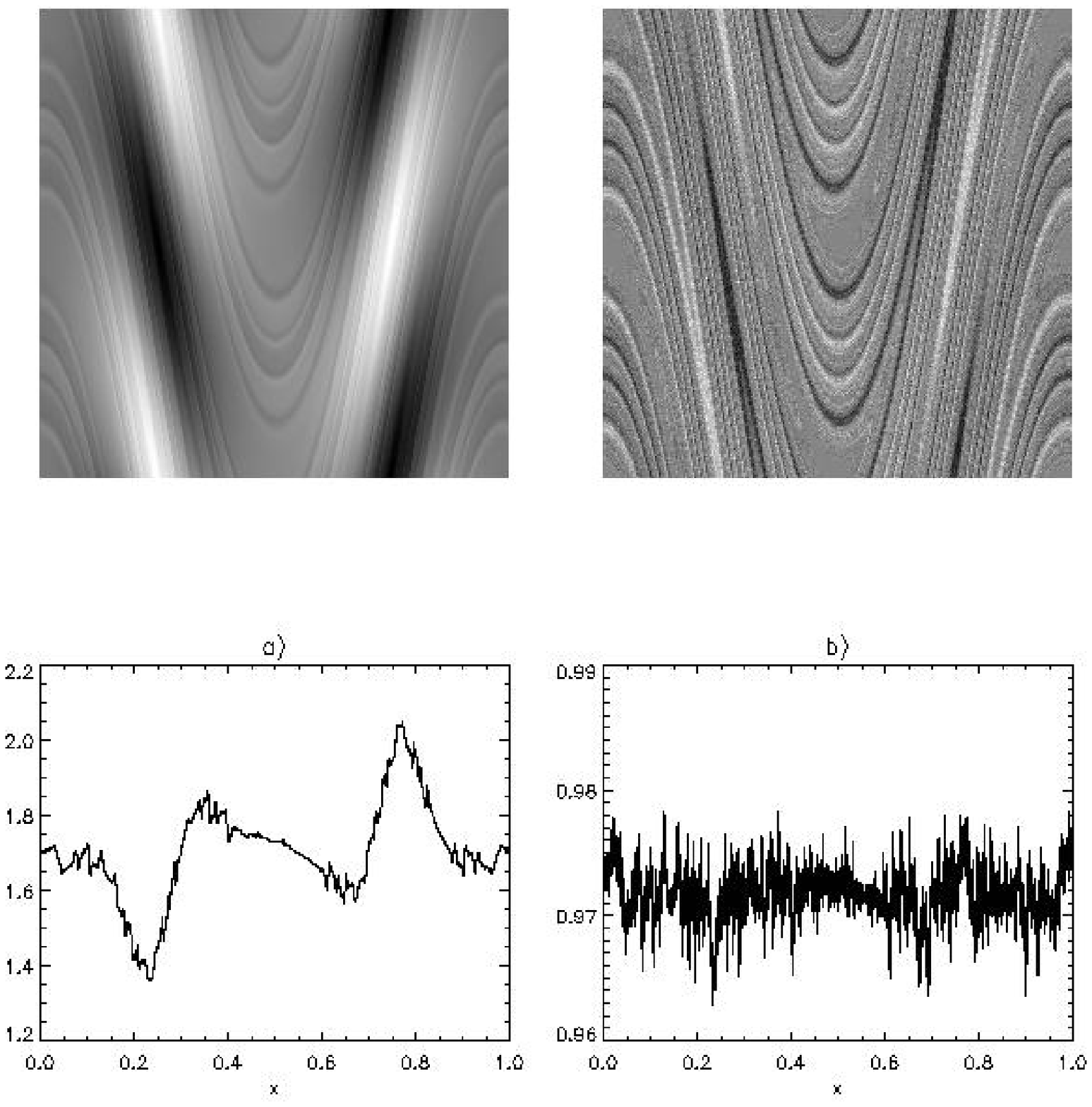,width=0.9\linewidth }
\vspace*{1 cm}
\caption{Carrying capacity (left) and zooplankton pattern (right) for $a=0.2$ and
$T=10$. Both are now rough, but with rather different smoothness
properties.}
\label{fig:2patsA02T10}
\end{figure}

By further reducing $T$, finally $C$ will become also filamental.
Confirmation of this prediction is given in
Figs.~\ref{fig:2patsA02T10} and \ref{fig:structA02T10} for $a=0.2$
and $T=10$. The measured value of $\lambda_C$ is $0.0380$. The
configurations of $C$ and $Z$ are displayed in
Fig.~\ref{fig:2patsA02T10} (the configuration of $P$ is very
similar to $Z$). The corresponding structure functions are in
Fig.~\ref{fig:structA02T10}, which shows that phytoplankton and
zooplankton have the same behavior at small scales, different from
the one of the carrying capacity. The predicted exponent for both
phyto and zooplankton is 0.16, very close to the fitting to the
two numerical curves $\alpha_{PZ}\approx 0.14$. The prediction
$\alpha_C\approx 0.85$ is more different from the observed
$\alpha_C \approx 0.75$, but it is still close.

Direct application of the arguments of Sect. \ref{sec:structure}
would indicate that the $P$ and $Z$ patterns are not smooth along
the direction of the filaments, since they would inherit there the
H\"{o}lder exponent of the source $C$. This conclusion is
incorrect, since the source is also filamental, and thus strongly
anisotropic. The direction of the filaments in $P$ and $Z$ is the
same as for $C$. The forcing $C$ is smooth along that direction,
and thus this is the behavior inherited by $P$ and $Z$ along the
filaments.

\begin{figure}
\epsfig{file=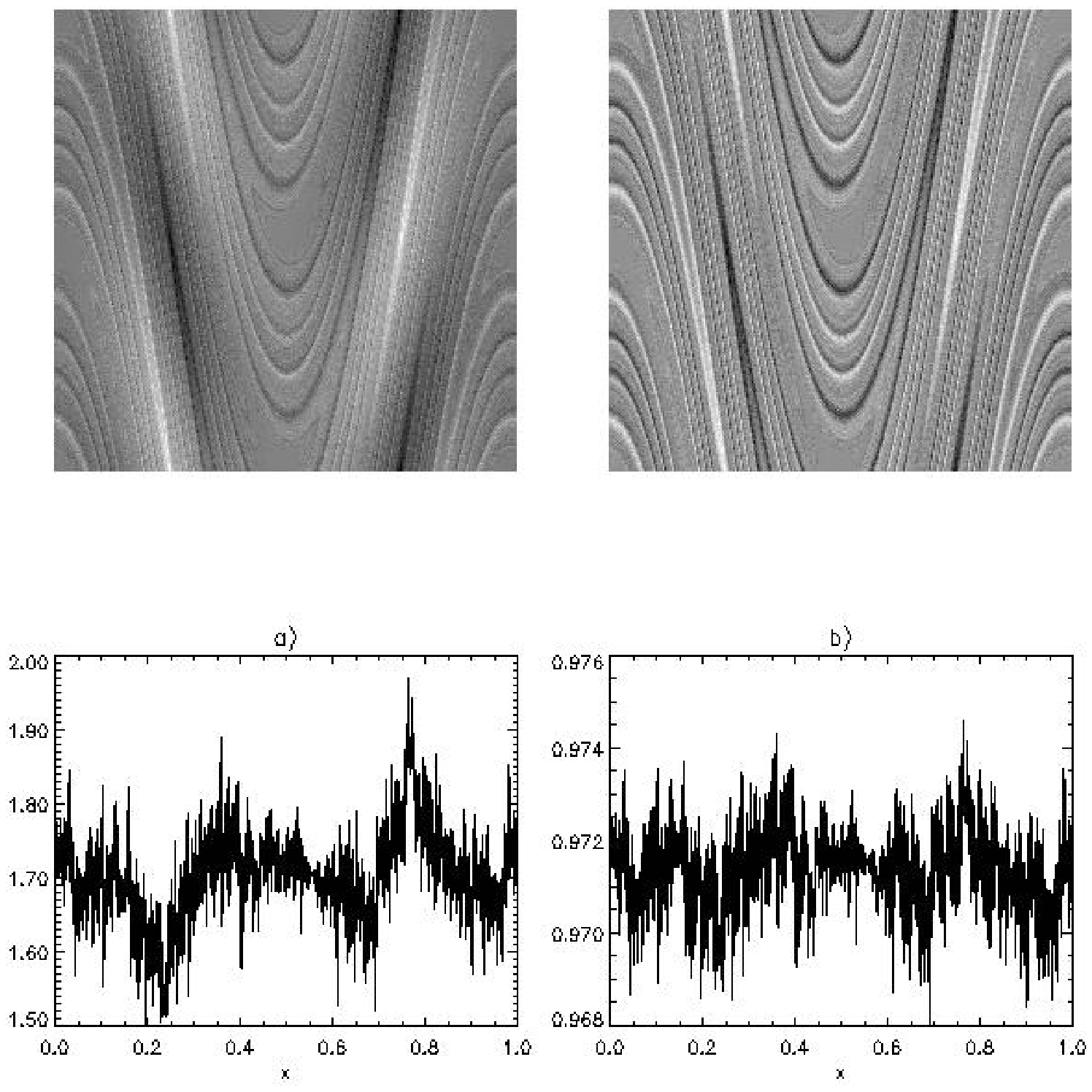,width=0.9\linewidth }
\vspace*{1 cm}
\caption{Carrying capacity (left) and zooplankton pattern (right) for $a=0.02$ and
$T=20$. Both display rough behavior of the same characteristics.}
\label{fig:2patsA0025T20}
\end{figure}

The second sequence of transitions one can find by decreasing $T$
occurs when $a<|\lambda_C|$: then the three fields, smooth at
large $T$, should become filamental at the same value of $T$, and
remain always with the same small-scale exponent, given by
$\alpha_C=\alpha_{PZ}=a/|\lambda_d|$.
Figure~\ref{fig:2patsA0025T20} confirms that the transition to
filamental behavior has occurred for $T=20$,
$a=0.025<|\lambda_C|\approx 0.0385$. The three structure
functions, displayed in Fig.~\ref{fig:structA0025T20}, have the
same fitting slope $\approx 0.20$, to be compared with the
prediction $\alpha_C=\alpha_{PZ}\approx 0.21$.

\section{Conclusions and open questions}
\label{sec:conclusions}

We have discussed the properties of the small-scale structure of
forced advected chemically or biologically active substances, and
show that they can be understood from the properties of models of
linearly decaying substances, as far as the decay rate is replaced
by the largest Lyapunov exponent of the chemical subsystem. We
have extended previous results to the case of non-smooth forcing.
The results have been applied to a three-component model of
plankton dynamics, which presents a particular asymmetric coupling
which requires special consideration. The morphological
transitions predicted by the theory are observed in the numerical
simulations. In addition, the numerical values of the predicted
scaling exponents for the first-order structure functions are very
close to the observed ones, despite of the approximations made.
This agreement would certainly deteriorate when considering
higher-order structure functions.

\begin{figure}
\epsfig{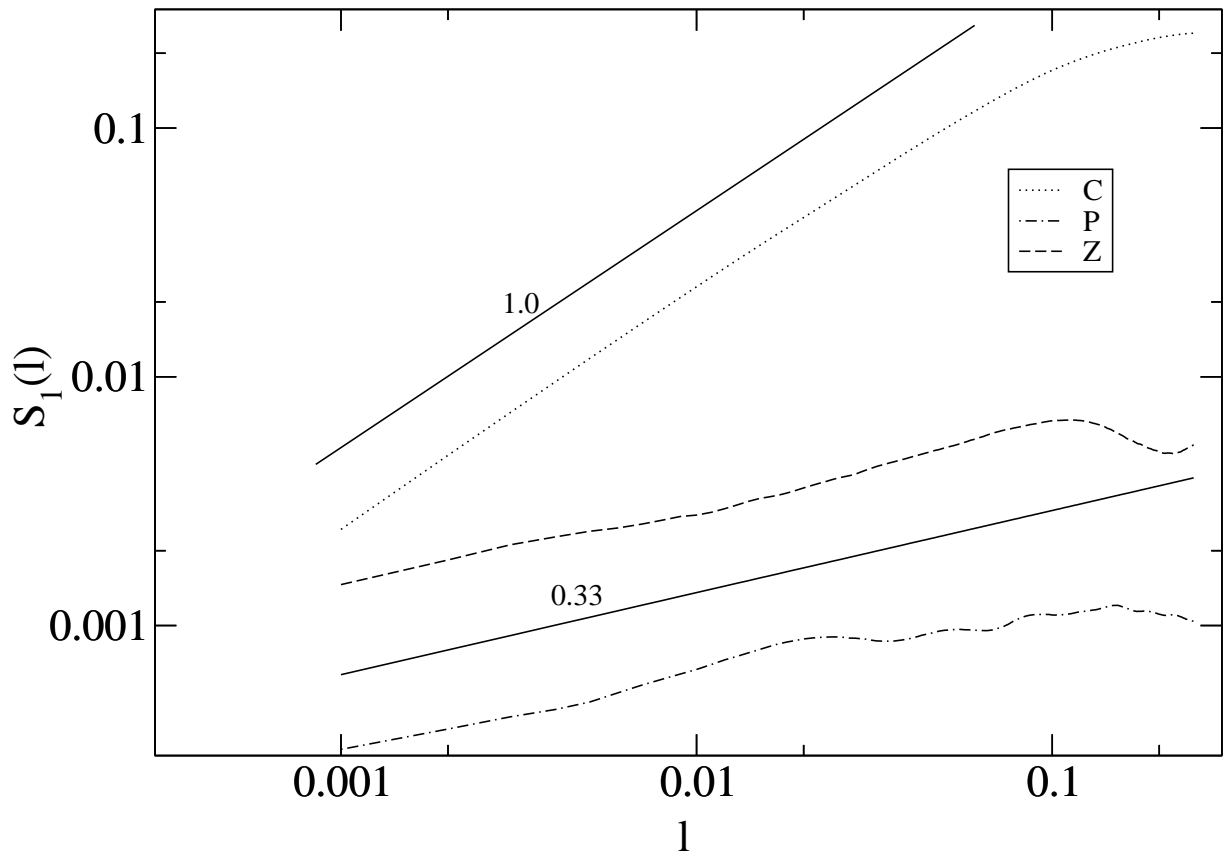}
\vspace*{1 cm}
\caption{First-order structure functions, $S_1^i(l)$, where $l\equiv \delta x$ is the
spatial separation, obtained by averaging over 50 transects in the
patterns of Fig.~\ref{fig:3patsA02T20}. The straight lines are the
theoretical predictions for the small scale behavior, and the
numbers above them the predicted slopes. }
\label{fig:structA02T20}
\end{figure}

\begin{figure}
\epsfig{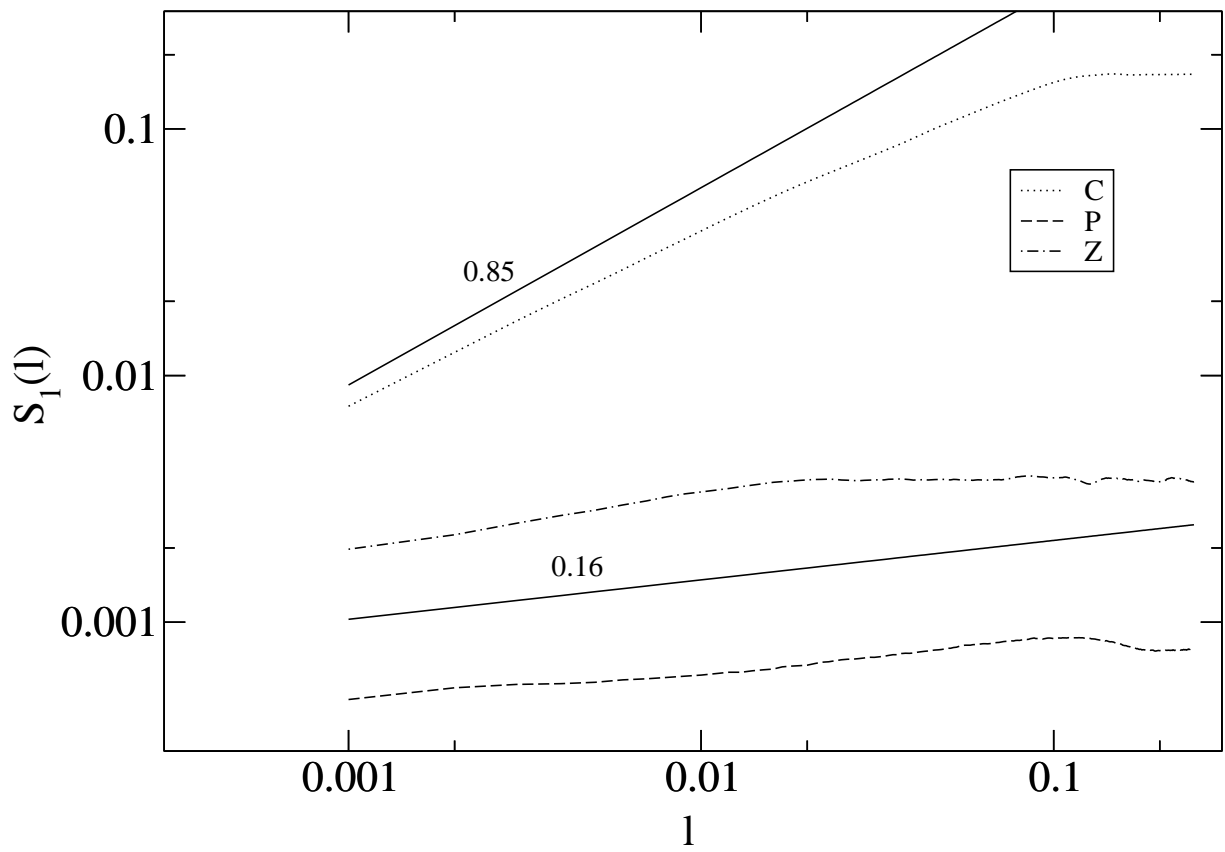}
\vspace*{1 cm}
\caption{Same as Fig.~\ref{fig:structA02T20}, but for the patterns in
Fig.~\ref{fig:2patsA02T10}.}
\label{fig:structA02T10}
\end{figure}

\begin{figure}
\epsfig{file=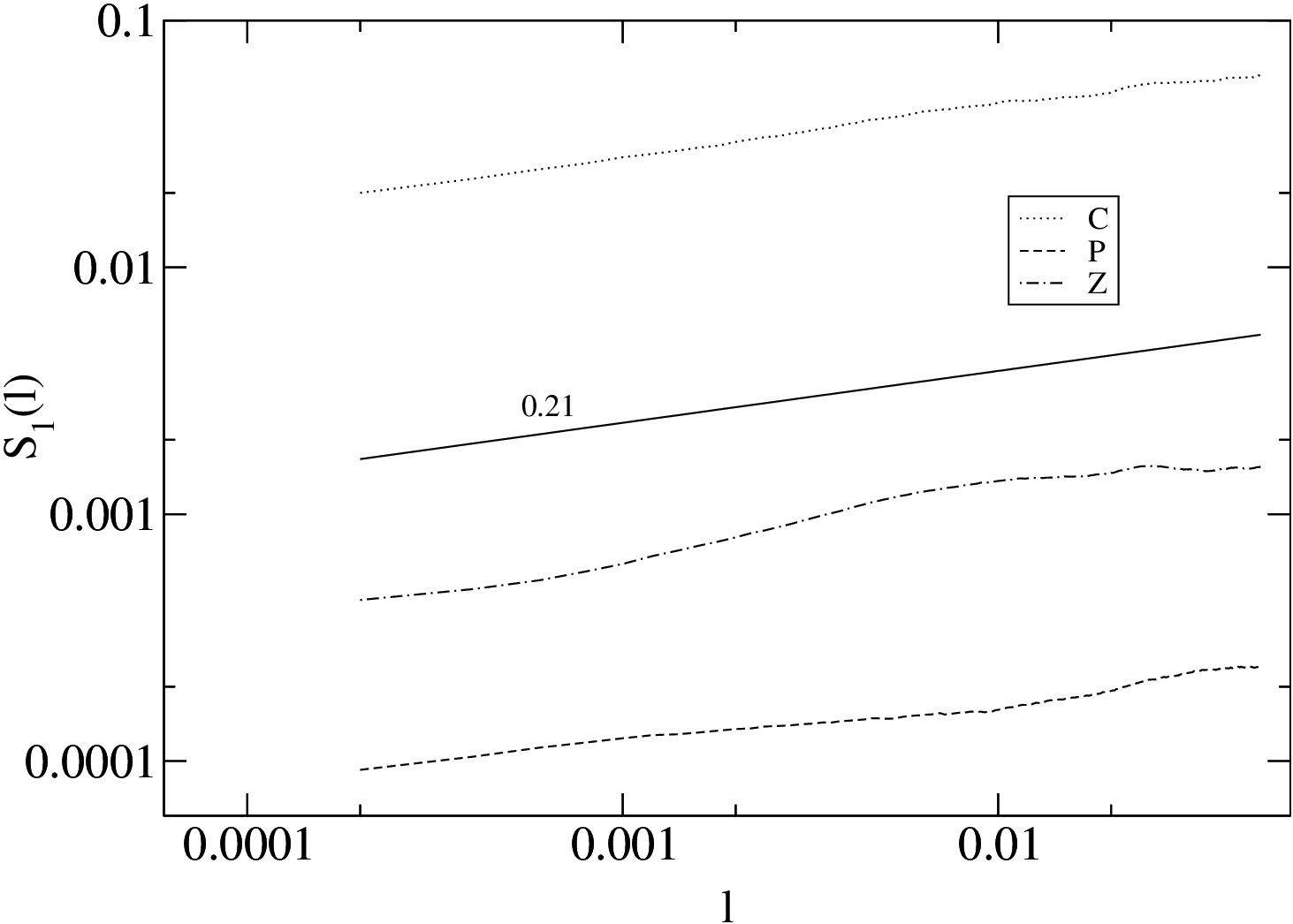,width=0.9\linewidth }
\vspace*{1 cm}
\caption{Same as Fig.~\ref{fig:structA02T20}, but for the patterns in
Fig.~\ref{fig:2patsA0025T20}.}
\label{fig:structA0025T20}
\end{figure}

An important point to remark is that our theory only applies to
the small scales of the advected patterns. Figures
\ref{fig:structA02T20}, \ref{fig:structA02T10} and
\ref{fig:structA0025T20} do not show too strong variations in
slope for different scales, but this is not guaranteed for any
model. This fact should be taken into account when applying the
results of this Paper to experimental situations and, specially,
to environmental flows. Constructing a theory valid beyond the
smallest scales remains an open challenge.

One of the consequences of our results is that, as long as one
restricts to the consideration of chaotic advection, and of the
small-scale structure, the only mechanism leading to differences
in the scaling properties of different interacting fields is the
presence of asymmetric couplings. This could be relevant to the
understanding of the differences that seem to appear in the
scaling behavior of different plankton species in the same flow
\cite{Abraham,Mackas79,Steele92}.

We have neglected diffusion in all our considerations. The
argument was that the effect of diffusion would be a smoothing of
the singularities below a diffusion scale, remaining unaltered the
behavior above that scale. Within the same framework used here,
this belief can be justified at least in the case of monocomponent
linear models \cite{diffusion}.

We have considered just the situation in which the largest
chemical Lyapunov exponent $\lambda_C$ remains negative even when
forced by stirring from the chaotic flow. The consideration of
positive Lyapunov exponents remains open, specially since we
expect diffusion to play a stronger role in the very singular
configurations that can be generated.

Finally, the consideration of multifractal behavior would be a
clear improvement of the theory. It seems straightforward to
include fluctuations in the Lyapunov exponents for the flow
subsystem in the same way as in \cite{PRE}. The result would be
that the structure function exponents get a corrections which
depend on the probability distribution of the finite-time Lyapunov
exponents. It seems more difficult to include the fluctuations of
the Lyapunov exponents of the chemical subsystem, since they are
not independent variables but depend on the statistics of the flow
exponents.

\begin{acknowledgements}
E.H-G acknowledges support from MCyT (Spain) projects BFM2000-1108
(CONOCE) and REN2001-0802-C02-01/MAR (IMAGEN). C.L. is supported
by the Spanish MECD.
\end{acknowledgements}

\newpage

\begin{thebibliography}{xx}
\bibitem{numerics} The spatial configurations displayed in the
figures have been obtained by integrating until a time $t$, large
enough, the chemical subsystem with Lagrangian trajectories
$\bX_t^0(\bx)$ that end at spatial points $\bx$ on a regular
lattice.
\end{thebibliography}
\end{document}